# A Novel Watermarking Approach for Protecting Image Integrity based on a Hybrid Security Technique

Ahmad M. Nagm
Electrical Engineering, Faculty of Engineering, Al Azhar University, Egypt

Mohamed Torky
Higher Institute of Computer Science and Information Systems-Culture & Science City- Egypt Member in Scientific Research Group in Egypt (SRGE)

Khaled Y. Youssef
Communications Engineering, Faculty of Space Technology, BSU University, Egypt

## ABSTRACT
Digital Photo images are everywhere around us in journals, on walls, and over the Internet. However we have to be conscious that seeing does not always imply reality. Photo images become a rich subject of manipulations due to the advanced digital cameras as well as photo editing software. Accordingly, image forgery is becoming much easier using the existing tools in terms of time and accuracy, and thus the forensics of detecting an image forgery case is becoming difficult and needs more and more time and techniques to prove the image originality especially as crime evidences and court related cases. In this paper, a framework with associated algorithms and methodologies is proposed to ensure the authenticity of the image and the integrity of the content in addition to protecting the photo image against forgery suspects. The framework depends on developing new generation of certified digital cameras that could produce authenticated and forgery-proof photos. The proposed methodology generates an irreversible hash integrity code from the image content based on color matrix calculations and steganography algorithms. The simulation results proved the capability of the proposed technique to detect image forgery cases in more than 16 scenarios of manipulation.

## 1. INTRODUCTION
In recently years, the manipulation of digital image more become easier, by using photo edit programs that may be embedded on the image capture devices as well[1], where Forgery detection aims to discriminate the digital images are original or false [2-3]. In this paper, a novel image security dynamic architecture framework is proposed for image authentication and originality identification techniques and related methodologies, algorithms and protocols that are applied on camera captured images [4]. The approach depends on implanting secret signs into RGB images that should indicate any unauthorized modification on the image under investigation [5]. The secret code generation depends mainly on two main parameter types, namely the image characteristics and capturing device identifier. On the other word, The general aim is detect images forgery by designing Algorithms to product secret code is called Secret Originality Identifier (SOI) have a dynamic properties and embed it into digital images with manner that is unnoticeable and in the same time must be distributed on every pixel of digital images [6], based on the CFA structure of the input mosaic image [7], where the captured images are passed on three filters and the output are three matrix values for every pixel that is represent the light intensity level coded at 8bit in 24bit system [8].

## 2. IMAGE AUTHENTICATION ARCHITECTURE: AN OVERVIEW
In this section, the architecture framework will be analyzed, explained and discussed together with the associated protocols, and algorithms. The concept of architecture is based on introducing new generations of digital cameras that is forgery-proof and their produced photos could be used as an official evidences of events specially the surveillance cameras used in strategic places and on streets. The philosophy behind is to enable the camera stamp the image using a photo authentication algorithm with a predefined hardcoded key that is specific to each camera (a key should be unique to cameras that is similar to the Ethernet MAC address concept in network cards). On the other side, a central verification server is proposed to apply a reverse algorithm and approve the originality of the image using two modes of operation namely the offline mode and the online mode. The server could be hosted by central authority globally or at country level through a governance framework and within a high grade security shell. In addition, a hash is generated [9-10] from the camera key and added to the photo image as a reference for the verification server to identify the corresponding key using secured mapping tables.

### 2.1 Offline Mode of Operation
In this mode, the camera is not a must to be IP or network camera, but it is essential to have a figure print of each camera that will be used through public photo authentication (PPA) protocol to check and verify that the photo is original and not manipulated using photo manipulate application. The camera should be equipped with a unique identity (serial could be identity the camera). In the offline operation mode, the user just verify an image under investigation regardless of the source and without any information on the author [11]. The user request the verification service directly from a front-end web service that enables back to back verification with the PAS server that should be located in a different DMZ zone for security purpose.

### 2.2 Online mode of operation
In this mode, the camera is an IP or network camera that should be equipped with camera MAC address and ID. The online mode of operation has two authentication modes:

#### 2.2.1 Public Authentication Mode
In this mode, the camera ID exists is an identifier on any registered image capture device that exist on both the equipment side (hardcoded on camera) as well as the authentication server register side known as Camera ID Register (CIDR) .The camera ID is thereafter transformed into photo ID using a convenient hashing algorithm which is the secret code word proposed to be embedded and hidden in the image post capture and is recognized as a forensic evidence for originality of image [12]. In this mode of authentication, the verification party will send the image under investigation to the Photo Authentication Server (PAS) that will in turn do the following:

- Extract the photo ID from the image.

- Query the CIDR on the photo ID to get the corresponding camera ID.

- Apply the technique standard hashing algorithm on the image using the camera ID and deduce the photo ID [13].

### 2.2.2 Private Authentication Mode

In the private authentication mode, the proposed framework reserve a space for private image communications for enterprises and closed groups. In this mode of authentication, the users can verify the image using a secret code that result from applying a hashing algorithm on the image. In this case, the algorithm will not use neither camera ID not the PPAP protocol but just an initial verification of the image within an organization using series of keys known for the organization. The disadvantage of this mode, that it could not be considered as an official model legally and thus cannot be used as an evidence proof for forensic techniques.

## 2.3 Public Photo Authentication Protocol

The protocol introduced to the aforementioned architecture is a novel protocol devised for this purpose named as "Public

Photo Authentication Protocol (PPAP)". The PPAP protocol is designed to enable the communication with the authentication server and accordingly the verification of the image for the different authentication modes discussed earlier. The PPAP protocol is governing the communication between the photo receiver side as a verification query point and the photo image authentication system as depicted in **Figure 1.**

## 2.4 Photo Authentication Algorithm

The photo authentication algorithm is the security algorithm concerned with integrity protection of the image in a sense that indicates also the originality of the image (i.e. the image being investigated is the image created by the first author authenticated digital camera). The algorithm target the protection of RGB images against forgery cases by implanting a generated secret originality identifier (SOI) into color captured images such that the SOI is distributed over the whole image according to algorithm [7]. The algorithm proposed consists of 6 main stages as depicted in **Figure 2**, post the photo image capture starting with selecting one of the three main color components as modifier component and another one as modified component [14-15]. Then the modifier component is ciphered using the AES algorithm [16-17] and the camera hard coded ID as security key to produce ciphered component.

Finally, a selected bit in the modifier component is replaced with the corresponding bit in the modified component [18-19-20]. Another version of the algorithm is proposed whereas a frequency transformation [21] step is added to extract the modifier bit from one of the frequency components of the image as explained later in simulation section.

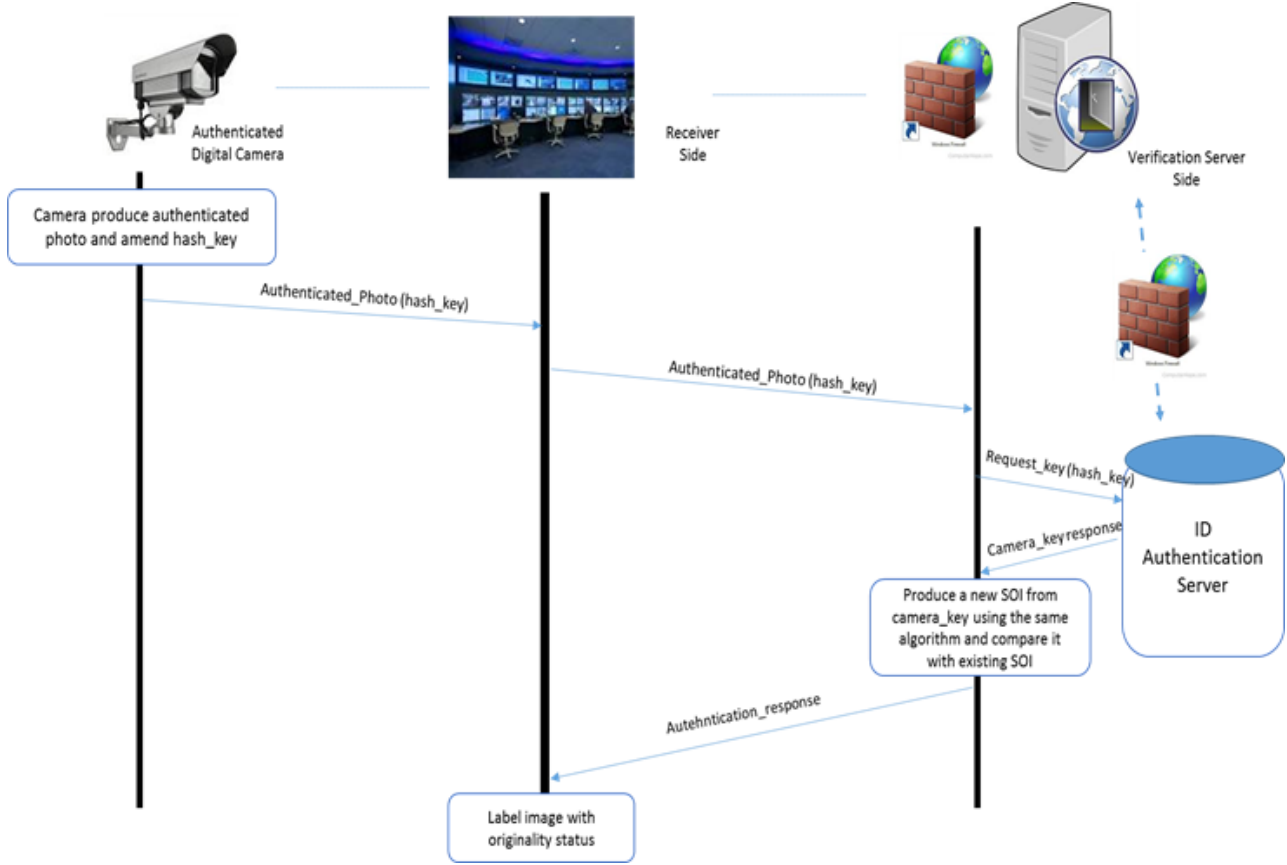

**Fig. 1 Public photo image Authentication protocol for online mode of operation.**

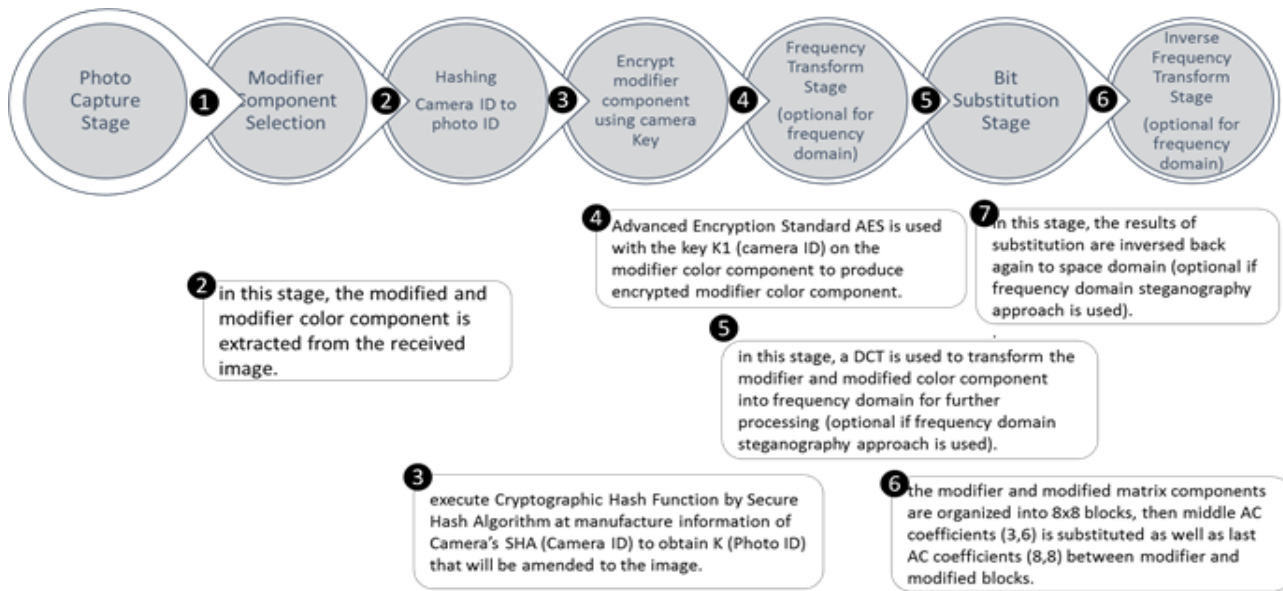

**Figure 2 Photo Authentication algorithm proposed for use in digital cameras and verification servers.**

**Table 1 six simulation scenarios grouped into two groups, Group A for frequency domain and Group B for space domain**

| Group | Scenario 1 | Scenario 2 | Scenario 3 |
|---|---|---|---|
| **Group A**<br><br>Originality steganography using single color modifier, In the frequency domain manipulation, captured image is transformed using DCT. | Steganography using Last AC Coefficient Mono Component for 8*8 block at color images. In this scenario, substitution process is performed at frequency domain after transform ciphered modifier and modified component, by Last AC Coefficient (8, 8) of every 8*8 block at color images. | Steganography using DC Coefficient Mono Component for 8*8 block at color images. In this scenario, substitution process is performed at frequency domain after transform ciphered modifier and modified component, by Last DC Coefficient (1, 1) of every 8*8 block at color images. | Steganography using First AC Coefficient Mono Component for 8*8 block at color images. In this scenario, substitution process is performed at frequency domain after transform ciphered modifier and modified component, by First AC Coefficient (1, 2) of every 8*8 block. |
| **Group B**<br><br>...lity Steganography will be executed using single color as a modifier component on space domain | substitution bit is the bit no. 4 in the modified color component (Red component) in this scenario, the forth bit in the ciphered modifier pixel component (blue color component) substitute the corresponding forth one in the unchanged modified pixel component (red color component) | the least significant bit in the ciphered modifier pixel component (blue color component) substitute the corresponding least significant one in the unchanged modified pixel component (red color component) | The most significant bit in the ciphered modifier pixel component (blue color component) substitute the corresponding most significant one in the unchanged modified pixel component (red color component) |

## 3. SIMULATION RESULTS

The simulation strategy is based on two main approaches, the first approach to assess the impact of the change on the space domain and frequency domain while the second approach is based on simulating an active attacks and assess the immunity of the system against several attacks attempts. The simulation tool used is MATLAB Version 8.5.0 with developed algorithms discussed in section 2 is applied on three images of diversified patterns and color spectrum on 6 scenarios both on space domain and frequency domain. On the space domain, the secret originality identifier (SOI) is deduced from the modifier RGB component (blue component for example) using the PPAP image authentication algorithm and implanted in the nth order bit in every byte of the modified RGB component (red component for example). On the other side, the change in frequency domain is simulated as explained in scenarios presented in Table 1. Table 2 depicts the application results of applying seven steganography scenarios on three images using the proposed photo authentication algorithm. Moreover Tables 3 show the averages of evaluation results of image quality measurements based on the proposed approach with respect to space domain. On the other hand Tables 4 show the averages of evaluation results of the image quality measurements with respect to frequency domain

## 4. DISCUSSION

Image authentication using a framework of protocol and algorithm in addition to a specific architecture is a subject that dictates standardization bodies adopting the idea and the significant need to apply this framework especially in the challenges introduced daily out of a fast moving digital world. The photo editing techniques are becoming accurate day after day and the capability to detect is becoming limited as well.

**Table 2 results of applying different Seven steganography techniques using the photo authentication algorithm**

| Experiment | Image 1:<br>Lina/220*220, 7.37KB, JPG | Image 2:<br>Baboon/267*266, 50.7KB, PNG | Image 3:<br>Pepper/512*512, 768KB, BMP |
|---|---|---|---|
| **Before SOI code insertion** | 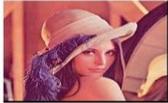 | 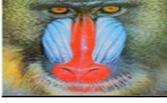 | 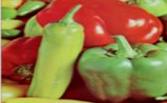 |
| **After SOI insertion (4-Bit)** | 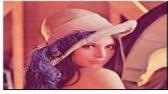 | 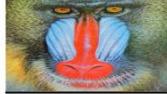 | 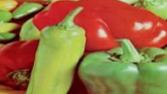 |
| **After SOI insertion (LSB)** | 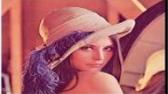 | 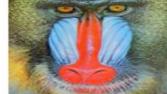 | 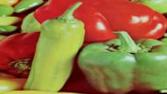 |
| **After SOI insertion (MSB)** | 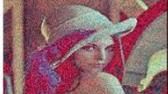 | 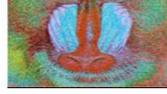 | 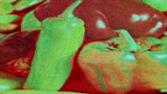 |
| **After SOI insertion (DC-Single)** | 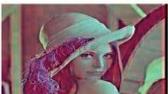 | 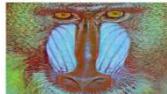 | 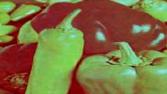 |
| **After SOI insertion (First AC-Single)** | 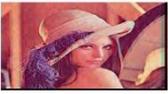 | 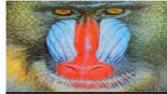 | 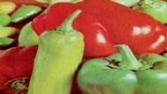 |
| **After SOI insertion (Last AC-Single)** | 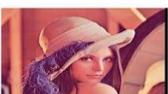 | 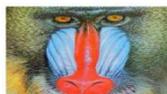 | 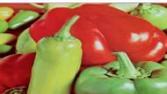 |
| **After SOI insertion (MID AC-Single)** | 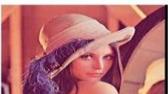 | 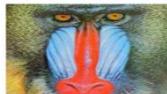 | 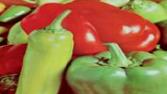 |

**Table 3 Averages of Image Quality Measurements of the Proposed Approach based on Space Domain for the three images : Lina, Baboon and Pepper**

| Index Type | | Space Domain | | |
|---|---|---|---|---|
| | | LSB SOI | 4th bit SOI | MSB SOI |
| **Mean Absolute Error (MAE)** | **Mean** | 0.0863 | 0.683466667 | 7.115966667 |
| | **Standard deviation** | 0.001276715 | 0.021050495 | 2.629197422 |
| **Mean Square Error (MSE)** | **Mean** | 0.167279268 | 10.70781038 | 2709.752752 |
| | **Standard deviation** | 0.001195837 | 0.024755073 | 9.424007186 |
| **Peak Signal to Noise Ratio (PSNR)** | **Mean** | 55.89645618 | 37.83380462 | 13.80152448 |
| | **Standard deviation** | 0.030983248 | 0.010038408 | 0.015116349 |
| **Structure Similarty Index (SSIM)** | **Mean** | 0.999835822 | 0.992097979 | 0.444948269 |
| | **Standard deviation** | 0.000104025 | 0.004225048 | 0.078718769 |
| **Universal Image Quality Index (UIQI)** | **Mean** | 0.999911464 | 0.994338272 | 0.102539898 |
| | **Standard deviation** | 2.94296E-05 | 0.001832541 | 0.12676949 |

**Table 4 Averages of Image Quality Measurements of the  Proposed Approach based on Frequancy Domain for the three images :
Lina, Baboon and Pepper**

| Index Type | | Frequency Domain | | | |
|---|---|---|---|---|---|
| | | DC-Single | First AC-Single | MID.AC-Single | Last AC-Single |
| Mean Absolute Error (MAE) | Mean | 3.317833333 | 1.773033333 | 0.9858 | 1.021666667 |
| | Standard deviation | 1.886497899 | 0.408246743 | 0.040771436 | 0.019185759 |
| Mean Square Error (MSE) | Mean | 1042.332816 | 94.23548403 | 26.43709775 | 28.56679574 |
| | Standard deviation | 382.9185717 | 53.99625537 | 1.599034166 | 1.512842663 |
| Peak Signal to Noise Ratio (PSNR) | Mean | 18.13173316 | 28.85073253 | 33.9140858 | 33.57617126 |
| | Standard deviation | 1.502398401 | 2.42878178 | 0.267253687 | 0.226694694 |
| Structure Similarty Index (SSIM) | Mean | 0.728924243 | 0.964457297 | 0.984098468 | 0.983545239 |
| | Standard deviation | 0.154627212 | 0.013697327 | 0.007493867 | 0.006745234 |
| Universal Image Quality Index (UIQI) | Mean | 0.483491679 | 0.953117649 | 0.986085273 | 0.98488924 |
| | Standard deviation | 0.111808485 | 0.020622757 | 0.005001492 | 0.005652002 |

However, the prosed study is handling the subject from three different perspective

- Perspective 1: definition of architecture and framework of algorithms & protocols

- Perspective 2: impact of architecture on image quality

- Perspective 3: success of the architecture and associated algorithm to detect forgery in any region within the image.

According to five measures : Mean Absolute Errors (MAE), Structure Similarity Index (SSIM), Universal Image Quality Index (UIQI), Mean Square Error (MSE), and Peak Single to Noise Ration (PSNR) in figures 3(a,b)  and figure 4 (a,b), it is obvious that space domain technique showed better performance than frequency domain techniques as mean square error reaches 0.1 for lowest significant bit substitution case in space domain versus 24 in the mid AC signal component in frequency domain. Generally, the space domain LSB is the best technique that could be used for image authentication algorithm. Another interesting finding is that the algorithm could be applied to any image type with very small impact on the quality of the image and thus the image is stamped with secret originality indicator in a seamless way. Moreover, evaluating the efficiency of the proposed algorithm in detecting image forgery clarified is 100% success to detect image  forgery using the LSB technique.

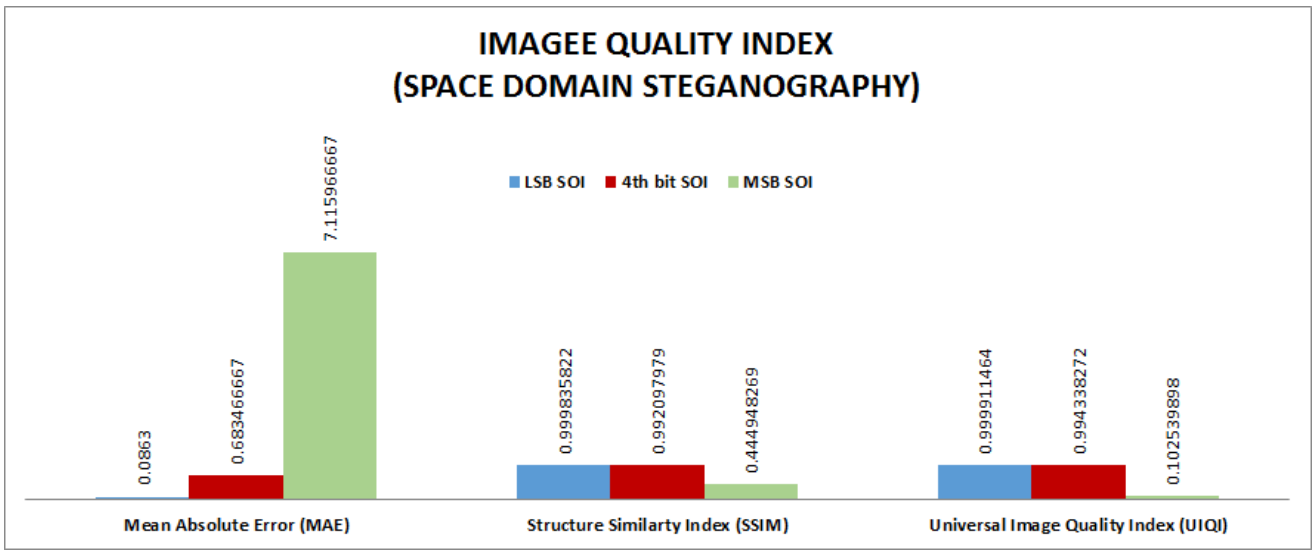

**(a)**

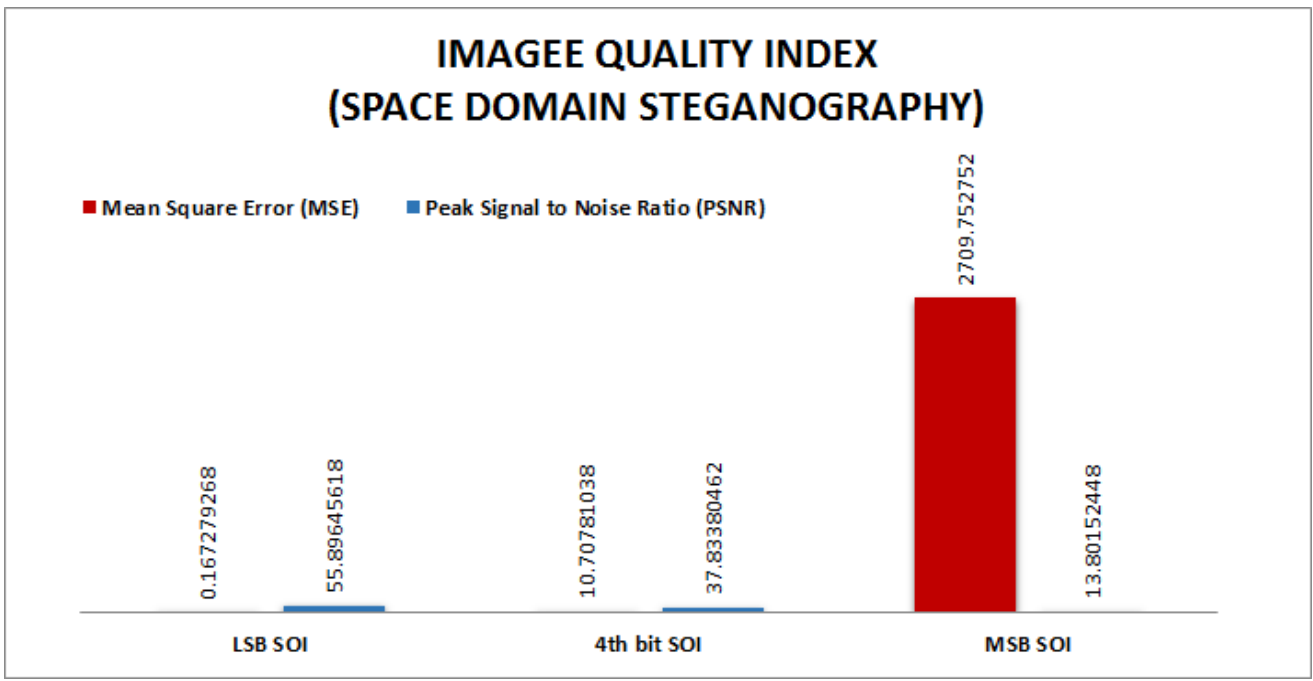

**(b)**

**Fig. 3 Space domain image quality indices: (a) MAE, SSIM, and UIQI results. (b) MSE, and PSNR results**

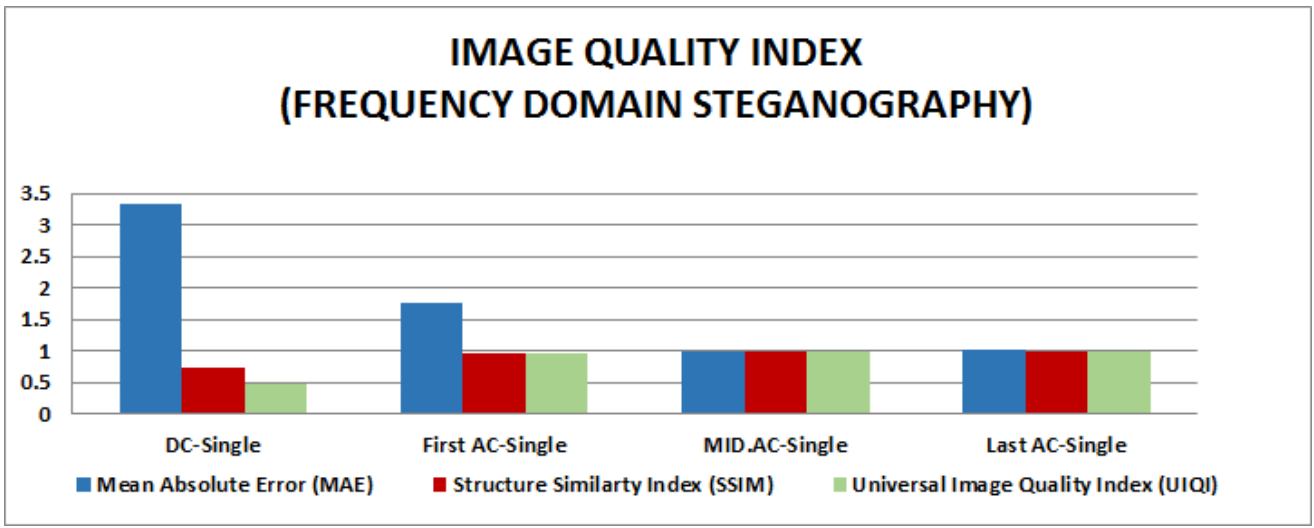

**(a)**

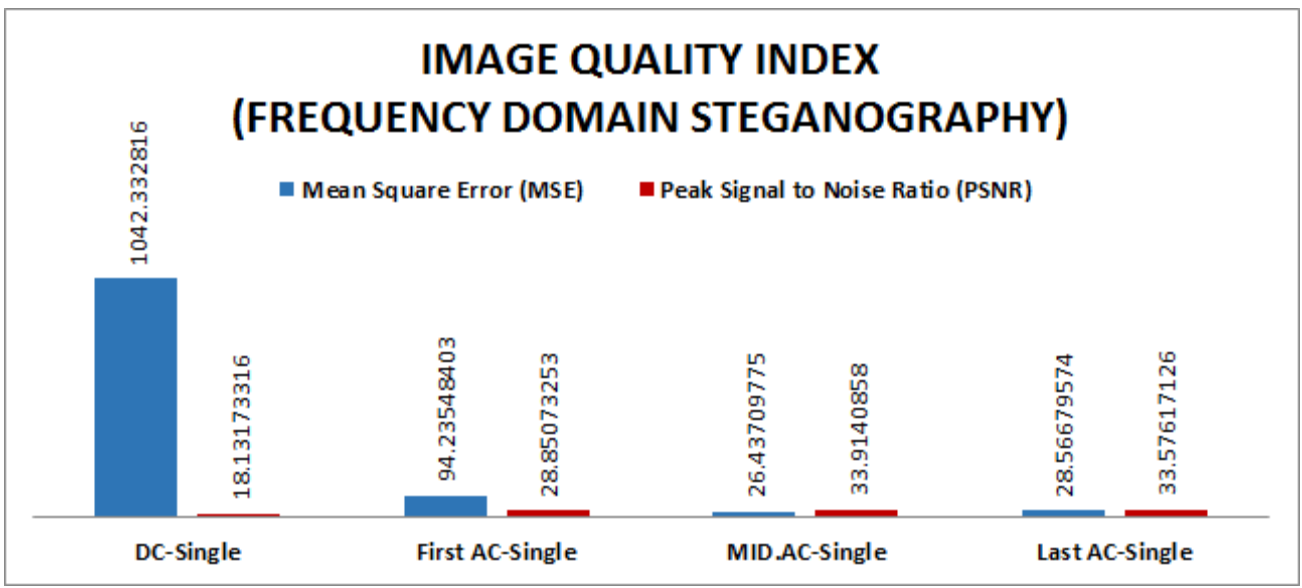

**(b)**

**Fig. 4 Frequency domain image quality indices: (a) MAE, SSIM, and UIQI results. (b) MSE, and PSNR results**

## 5. CONCLUSION

There is a significant need for photo image authentication architecture that should be adopted by standardization bodies to ensure the originality of digital photos against forgery cases especially with the huge advances in technology of editing. The proposed framework of architecture, protocols and algorithms is proved to show accurate results on detecting the cases of forgery with minimal impact on image quality (0.01% on the average) regardless of the details or the spectrum of the image.

In general, the space domain technique showed better performance than frequency domain techniques as mean square error reaches 0.1 for lowest significant bit substitution case in space domain versus 24 in the mid AC signal component in frequency domain. Generally, the space domain LSB is the best technique that could be used for image authentication algorithm.

## 6. REFERENCES

[1] A. Cheddad, J. Condell, K. Curran, and P. Mc Kevitt, "Digital image steganography: Survey and analysis of current methods," Signal processing, vol. 90, no. 3, pp. 727–752, (2010).

[2] H Farid, A survey of image forgery detection. IEEE Signal Process. Mag. 2(26), pp. 16–25, (2009).

[3] B Mahdian, S Saic, A bibliography on blind methods for identifying image forgery. Signal Process. Image Commun. 25(6), pp. 389–399, (2010).

[4] I. Cox, M. Miller, J. Bloom, J. Fridrich, and T. Kalker, Digital watermarking and steganography. Morgan Kaufmann, 2007.

[5] JG Han, TH Park, YH Moon, IK Eom, Efficient Markov feature extraction method for image splicing detection using maximization and threshold expansion. J. Electron. Imaging. 25(2), pp. 23-31 (2016).

[6] WC Hu, WH Chen, DY Huang, CY Yang, Effective

image forgery detection of tampered foreground or background image based on image watermarking and alpha mattes. Multimed. Tools Appl. 75(6), pp. 3495–3516, (2017).

[7] CH Choi, HY Lee, HK Lee, Estimation of color modification in digital images by CFA pattern changes. Forensic Sci. Int. 226, pp. 94 –105, (2013).

[8] P Ferrara, T Bianchi, A De Rosa, A Piva, Image forgery localization via finegrained analysis of CFA artifacts. IEEE Trans. Inf. Forensics Secur. 7(5), pp. 1566– 1577, (2012).

[9] Che-Yen, Wen, and Yang Kun-ta. "Image authentication for digital image evidence." Forensic Science Journal, 2006: 1-11.

[10] Scientific Working Group on Imaging Technology (SWGIT) , Draft Recommendations and Guidelines for the Use of Digital Image Processing in the Criminal Justice System, Version 1.1 , February 2001.

[11] J Lukˊaˊs, J Fridrich, M Goljan, Digital Camera Identification from Sensor Pattern Noise. IEEE. T. INF. Foren. Sec. 1(2), pp. 205- 214, (2006).

[12] I Amerini, L Ballan, R Caldelli, A Del Bimbo, G Serra, ASIFT-based forensic method for copy move attack and transformation recovery. IEEE. Trans. Inf. Forensics Secur.6 (3), pp. 1099–1110, (2011).

[13] S Bayram, HT Sencar, N Memon, Classification of digital camera-models based on demosaicing artifacts. Digit. Invest. 5, pp. 49 –59, (2008).

[14] H Cao, AC Kot, Accurate detection of demosaicing regularity for digital image forensics. IEEE Trans. Inf. Forensics Secur. 4(4), pp. 899–910, (2009).

[15] AC Gallagher, TH Chen, Image authentication by detecting traces of demosaicing. Proceedings of IEEE Computer Society Conference on Computer Vision and Pattern Recognition Workshops, pp. 1–8, (2008).

[16] D. R. Stinson, Cryptography: theory and practice. CRC press, 2005.

[17] D. B. A. M. D. E. I. M. Yousif Elfatih Yousif, "Review on Comparative Study of Various Cryptography Algorithm," International Journal of Advanced Research in Computer Science and Software Engineering(IJARCSSE), vol. 5, no. 4, pp. 51-55, 2015.

[18] Suresh Kumar, Ganesh Singh, Tarun Kumar(2013), "Hiding the Text Messages of Variable Size using Encryption and Decryption Algorithms in Image Steganoglaphy", International Journal of Computer Applications (0975 – 8887) Volume 61– No.6, January 2013.

[19] M.RAJKAMAL and B.S.E. ZORAIDA, (2014) "Image and Text Hiding using RSA & Blowfish Algorithms with Hash-Lsb Technique", International Journal of Innovative Science, Engineering & Technology, Vol. 1 Issue 6, August 2014.

[20] G.R.Manjula and AjitDanti, (2015) "A NOVEL HASH BASED LEAST SIGNIFICANT BIT (2-3-3) IMAGE STEGANOGRAPHY IN SPATIAL DOMAIN", International Journal of Security, Privacy and Trust Management, Vol 4, No 1, February 2015.

[21] Z Lin, J He, X Tang, CK Tang, Fast, automatic and fine-grained tampered JPEG image detection via DCT coefficient analysis. Pattern Recognition 42(11), pp. 2492–2501, (2009)